\documentclass[12pt]{article}
\usepackage{epsfig,amsfonts,amssymb}
\usepackage{amsmath}
\usepackage[margin=2.5cm]{geometry}
\usepackage[colorlinks=true,urlcolor=blue,linkcolor=blue]{hyperref}
\usepackage{cite}
\input epsf.sty
\usepackage{cite}

\def\ZZZ{{\hbox{ Z\kern-1.6mm Z}}}
\def\RRR{{\hbox{ R\kern-2.4mm R}}}
\def\CCC{{\hbox{ C\kern-2.0mm C}}}
\def\zzz{{\hbox{z\kern-1mm z}}}

\newcommand{\vt}{\vartheta}

\newcommand{\qeq}{{\hbox{=\kern-2.3mm ? \kern.5mm }}}
\renewcommand{\qeq}{=}

\newcommand{\ve}{\varepsilon}

\newcommand{\MM}{{\cal M}}

\newcommand{\WW}{{\cal W}}

\newcommand{\wt}{\widetilde}
\newcommand{\wh}{\widehat}

\newcommand{\NN}{{\cal N}}

\newcommand{\be}{\begin{equation}}
\newcommand{\ee}{\end{equation}}
\newcommand{\ben}{\begin{eqnarray}\displaystyle}
\newcommand{\een}{\end{eqnarray}}

\newcommand{\refb}[1]{(\ref{#1})}
\newcommand{\p}{\partial}
\newcommand{\sectiono}[1]{\section{#1}\setcounter{equation}{0}}

\def\one{{\hbox{ 1\kern-.8mm l}}}
\def\zero{{\hbox{ 0\kern-1.5mm 0}}}

\begin{document}

\baselineskip 24pt

\begin{center}
{\Large \bf  BPS State Counting in N=8 Supersymmetric String Theory for
Pure D-brane Configurations}

\end{center}

\vskip .6cm
\medskip

\vspace*{4.0ex}

\baselineskip=18pt

\centerline{\large \rm Abhishek Chowdhury, Richard S.~Garavuso, 
Swapnamay Mondal, Ashoke Sen}

\vspace*{4.0ex}

\centerline{\large \it Harish-Chandra Research Institute}
\centerline{\large \it  Chhatnag Road, Jhusi,
Allahabad 211019, India}

\vspace*{1.0ex}
\centerline{\small E-mail:  abhishek,garavuso,swapno,sen@hri.res.in}

\vspace*{5.0ex}

\centerline{\bf Abstract} \bigskip

Exact results for the BPS index are known for a class of BPS dyons in type II string theory
compactified on a six dimensional torus. In this paper we set up the problem of counting the same
BPS states in a duality frame in which the states carry only Ramond-Ramond charges.
We explicitly count the number of states carrying the lowest possible charges and find agreement
with the result obtained in other duality frames. Furthermore, 
we find that after factoring out the supermultiplet structure, each of these states 
carry zero angular momentum. This is in agreement
with the prediction obtained from a representation of these states as supersymmetric black
holes.

\vfill \eject

\baselineskip=18pt

\tableofcontents

\sectiono{Introduction}  \label{s1}

Understanding the microscopic origin of Bekenstein-Hawking entropy is one of the important
problems in any theory of quantum gravity, and in particular in string theory. In recent years
there has been considerable progress towards this direction, including precision counting of
microscopic states in certain string theories with 16 or more unbroken 
supersymmetries\cite{9607026,0510147,0609109,0802.1556,0803.2692,9903163,0506151}.
One of these theories is type IIA or IIB compactified on a six dimensional
torus. In this theory, for certain configurations carrying a combination of Kaluza-Klein (KK) monopole
charge, momentum along one of the circles of the torus and D-brane wrapping charges along
some of the cycles of the torus, one can carry out the exact counting for the number of microscopic
BPS states\cite{0506151}. On the other hand, for large charges this system can be described by a 
supersymmetric
black hole with a finite area event horizon. Thus, by comparing the logarithm of the number of 
microstates with the Bekenstein-Hawking entropy of the corresponding black hole, one can 
verify the equality of the macroscopic and microscopic entropy of the black hole.

Although the counting of microscopic states was carried out for a specific system of KK
monopoles and D-branes carrying momentum along a compact circle, using duality symmetry
we can map it to other systems. In particular it is possible to map this configuration to a system
that contains only D-brane charges. Duality symmetry predicts that the 
BPS index of this system computed from
microscopic counting should give the same result as the original system to which
it is dual. Nevertheless, it is of some interest to count the number of microscopic states of the new
system directly. At the least, this will provide us with another non-trivial test of duality symmetry
which, although has been tested in many ways, has not been proven. Another motivation for this
is that 
by learning how to count states of pure D-brane systems in type
II string theory on $T^6$ we may eventually gain some insight into similar counting for D-branes
wrapped on various cycles of Calabi-Yau manifolds.
Indeed,
for type II compactification on Calabi-Yau manifolds, all charges are associated with 
D-branes wrapped on various cycles of Calabi-Yau manifolds as there are no non-contractible
circles and hence
no momentum, KK monopole charges or winding numbers of fundamental strings and 
NS 5-branes. Earlier attempts to count states of pure D-brane systems describing a black hole
can be found in \cite{0509168,0607010,0702146}.

Another reason for studying representations of black holes as pure D-brane systems is as follows.
One knows on general grounds that supersymmetric black holes in 3+1 dimensions describe
an ensemble of states each of which carries strictly zero angular 
momentum\cite{0903.1477,1009.3226} after factoring out
the fermion zero mode contribution whose quantization 
generates the supermultiplet structure. 
This leads to many non-trivial conjectures about the sign of the index of supersymmetric black holes
which have been verified explicitly\cite{1008.4209,1208.3476}.
However, in
microscopic counting of the same system, one often finds BPS states carrying non-zero angular
momentum. This does not represent a contradiction between microscopic and macroscopic
results, since only the index, and not the detailed information about angular momentum, is
protected as we go from the weak coupling regime where microscopic calculation is valid, to the 
strong coupling regime where the black hole computation is valid. Nevertheless, one could
ask if there is a duality frame in which the detailed information about the angular momentum  in the microscopic and macroscopic descriptions matches.
Since in the macroscopic description all black
holes carry zero angular momentum, in the microscopic description this will demand that all
states are singlets under the SU(2) rotation group. 
Recent analysis of some microstates of $\NN = 2$ 
supersymmetric black holes revealed that when we describe them as D-branes wrapped on certain internal cycles of Calabi-Yau manifolds we indeed get exactly zero angular momentum for the microstates of single centered black holes\cite{1205.5023,1205.6511,1207.0821}.
Assuming this to be a general phenomenon led to the conjectured Coulomb branch formula
for computing the spectrum of quiver quantum mechanics and of general systems of 
multicentered black holes\cite{1103.1887,1207.2230,1302.5498}.

Now in $\NN=2$ supersymmetric string theories, the  above
analysis is made complicated due to the fact that 
the index receives a contribution from both single and
multi-centered black holes. Since the latter do not necessarily carry zero angular momentum, 
we need to carefully subtract the contribution from
multi-centered black holes before we can verify that D-brane microstates representing single
centered black holes carry zero angular momentum. 
This can be 
done\cite{0807.4556,1103.1887}, and was used in the analysis of 
\cite{1103.1887,1207.2230,1302.5498}. 
However, in type II string theory on $T^6$, which has $\NN=8$ supersymmetry, 
the multi-centered black holes do not
contribute to the index, and hence we expect that only single centered black holes will
survive at a generic point in the moduli space of the theory\cite{0803.1014}.
Generalization of the observations in $\NN=2$ supersymmetric string theories made
above would then suggest that
representing 
a supersymmetric black hole in type II on $T^6$
as a system carrying only
RR charges associated with various D-brane sources, we may get a system whose microstates
would have strictly zero angular momentum after factoring out the goldstino fermion modes
whose quantization generates the supermultiplet structure. Now, after factoring out these 
fermionic zero modes and the bosonic zero modes
associated with various translational symmetries, the BPS states of the D-brane system 
correspond to the cohomology of the moduli space of classical
solutions of the world-line theory of the system, 
with the space-time rotation group acting as the Lefshetz SU(2) action on the 
cohomology\cite{9907100,1205.5023}.
This shows that
in order to get only zero angular momentum states, all states must come from the middle
cohomology. Since any compact manifold has a non-trivial 0-form and a top form, the only way
that a manifold can have only middle cohomology is if it becomes zero dimensional, i.e.\ a collection
of points.\footnote{We thank Boris Pioline for discussion on this point.} 
Verification of this conjecture is another motivation for our analysis.

In this paper we shall analyze a pure D-brane system in type II theory 
compactified on $T^6$ that is dual to the system for which
the microscopic result is known, and test the result by direct computation of the microscopic
index of the D-brane system. We introduce the system in \S\ref{ssystem}, and derive its
world-line theory for the lowest possible values of the charges 
in \S\ref{s2}. In \S\ref{s3} we explicitly count the index of supersymmetric 
states of this system. 
This is shown to reduce to counting the number of independent solutions of a
set of polynomial equations -- a problem that can be easily solved. 
We find that the solution contains a set of isolated points provided we work at a generic
point in the moduli space of the theory parametrized by constant background values of
the metric and 2-form fields along the internal torus. 
Hence, at least in this example, the microstates carry strictly
zero angular momentum in agreement with the macroscopic results.
In \S\ref{sb} we briefly discuss possible generalization of our analysis to cases where
we 
replace each D-brane of the system described in \S\ref{ssystem} by a stack of
parallel D-branes. We conclude with a discussion of our results in \S\ref{sconc}.
In appendix \ref{sa}
we derive the 
relation between some of the parameters of the D-brane world-volume theory and the background
values of the metric and 2-form field along $T^6$. In appendix \ref{sc} we describe the chain
of dualities that relate the system under consideration to the system analyzed in 
\cite{0506151}. Finally, in appendix \ref{sd} we give explicit solutions to the polynomial equations
which appear in the analysis of \S\ref{s3}.

\sectiono{The system} \label{ssystem}

Let us consider for definiteness a type IIA string theory on $T^6$ labelled by the coordinates
$x^4,\ldots, x^9$ and in this theory we take a system
containing $N_1$ D2-branes wrapped along
4-5 directions, $N_2$ D2-branes wrapped along 6-7 directions, $N_3$ D2-branes wrapped
along 8-9 directions,
$N_4$ D6-branes wrapped along 4-5-6-7-8-9 directions and $N_5$ D4-branes along
6-7-8-9 directions.
By a series of duality transformations  
described in appendix \ref{sc},
this configuration is related to a system 
of $N_1$ KK monopoles associated with the 4-direction, $-N_2$ units of momentum along the
5-direction, $N_3$ D1-branes along the 5-direction, $N_4$ D5-branes along 5-6-7-8-9 directions
and $-N_5$ units of momentum along the 4-direction.
The microscopic index of this system was computed explicitly in \cite{0506151}
for $N_1=1$. By a further series of
U-duality transformations reviewed {\it e.g.} in \cite{0708.1270}, this system can be mapped to a
system in type IIA string theory on $T^6$ with only NS-NS sector charges,  
containing $-N_2$ units of momentum along the 5-direction, $N_1$ fundamental strings
wound along the 5-direction, $N_4$ KK monopoles associated with the 4-direction, $-N_3$
NS 5-branes wrapped along 5-6-7-8-9 directions and $N_5$ NS 5-branes along
4-6-7-8-9 directions. In the notation of 
\cite{0708.1270},  the electric charge vector $Q$ and magnetic charge vector $P$ 
for this state in the latter description are represented as  
\be
Q=\begin{pmatrix}0\cr -N_2 \cr 0\cr -N_1\end{pmatrix}, \quad P = \begin{pmatrix}
N_3\cr N_5\cr  N_4\cr 0\end{pmatrix}\, .
\ee 
The T-duality invariant 
inner product matrix between charges
was given by $\begin{pmatrix}0 & I_2\cr I_2 & 0\end{pmatrix}$.
With this we get
\be
Q^2 =2\, N_1 N_2, \quad P^2 = 2\, N_3 N_4, \quad Q\cdot P=-N_1 N_5\, .
\ee
We also define
\ben
\ell_1 &=& \gcd\{Q_i P_j - Q_j P_i\} =
\gcd\{ N_1 N_3, N_1 N_4, N_2 N_3, N_2 N_4, N_5 N_1\}, \nonumber \\
\ell_2 &=& \gcd\{Q^2/2, P^2/2, Q\cdot P\} =
\gcd\{ N_1 N_2, N_3 N_4, N_1 N_5\}\, .
\een
We shall consider configurations for which
\be 
\gcd\{\ell_1, \ell_2\}=1, \quad i.e.\ \quad 
\gcd\{ N_1 N_3, N_1 N_4, N_2 N_3, N_2 N_4, N_1 N_2, N_3 N_4, N_1  N_5\}=1 \, .
\ee
In this case, following \cite{0702150,0712.0043} one can show that there is a further
series of duality transformations that map this system to one with
$N_1=1$\cite{0804.0651} for which the 
microscopic index is known from the analysis of
\cite{0506151}. Expressed in terms of the more general set of variables $(N_1,\cdots,N_5)$,
the result for the BPS index for this system, which in this case corresponds to the
14-th helicity supertrace  $B_{14}$, takes the form\cite{0908.0039}
\be
B_{14} = (-1)^{Q\cdot P+1} \sum_{s|\ell_1 \ell_2} s\, \wh c(\Delta/s^2) \, , \quad \Delta \equiv 
Q^2 P^2 - (Q\cdot P)^2 = 4\, N_1 N_2 N_3 N_4 - (N_1 N_5)^2
\, ,
\ee
where $\wh c(u)$ is defined through the 
relation\cite{9903163,0506151}
\be \label{ek6.5}
-\vt_1(z|\tau)^2 \, \eta(\tau)^{-6} \equiv \sum_{k,l} \wh c(4k-l^2)\, 
e^{2\pi i (k\tau+l z)}\, .
\ee
$\vt_1(z|\tau)$ and $\eta(\tau)$ are respectively the odd Jacobi
theta function and the Dedekind eta function. 

In this paper we shall analyze the simplest of these configurations with 
\be \label{echarge}
N_1=N_2=N_3=N_4=1, \quad N_5=0\, .
\ee
For this, \refb{ek6.5} predicts 
\be \label{eb14}
B_{14} = 12\, .
\ee
We shall verify this by direct counting of microstates of the D-brane
system.

\sectiono{The low energy dynamics of the D-brane system} \label{s2}

The combined system of four D-branes that we have introduced in \S\ref{ssystem}
with the choice of $N_i$'s given in \refb{echarge}
preserves 4 out of the 32 supersymmetries. This
is equivalent to $\NN=1$ supersymmetry in 3+1 dimensions.
Since we are dealing with a quantum mechanical
system, we can effectively regard this as an $\NN=1$ supersymmetric theory
in 3+1 dimensions, dimensionally reduced to 0+1 dimensions. Thus we can
can use the $\NN=1$ superfield
formalism, but ignore all spatial derivatives and integration over spatial directions
while writing the action. We shall follow the normalization conventions
of \cite{9701069} in constructing this
action.

Since the four D-branes we have are
related to each other by T-duality, each  of them individually has the same low energy theory
given by the dimensional reduction of $\NN=4$ supersymmetric U(1) gauge theory
from 3+1 to 0+1 dimensions.  
We begin with one of the four different D-branes. In the language
of $\NN=1$ supersymmetry in 3+1 dimensions, each D-brane has one U(1) 
vector superfield $V$ and
three chiral superfields $\Phi_1,\Phi_2, \Phi_3$. 
A vector multiplet, after dimensional reduction to 0+1 dimensions, 
has three scalars corresponding to three spatial components of the gauge field 
and a gauge field $A_0$. We can use the gauge $A_0=0$ and interpret the three scalars as
the coordinates giving the location of the D-brane along the three non-compact directions.
We shall denote these three scalars by $X_1, X_2, X_3$. 
The three chiral multiples $\{\Phi_i\}$ contains three complex scalars
$\{\Phi_i\}$.\footnote{Following usual notation, we shall use the same symbol to denote a
superfield and its scalar component.}
These complex scalars give the coordinates or Wilson lines along $x^4+i x^5$, $x^6+i x^7$ and
$x^8+i x^9$ directions respectively. 
For example, for the D6-brane all three complex scalars
correspond to Wilson lines, while for the D2-brane wrapped along the 
4-5 directions, $\Phi_1$ corresponds
to a Wilson line along $x^4+ix^5$ but $\Phi_2$ and $\Phi_3$ correspond to positions of the brane
along $x^6+ix^7$ and $x^8+i x^9$ respectively. Finally, we shall use a superscript $(k)$ to
label the four different D-branes, with $k=1,2,3$ corresponding to D2-branes wrapped along the
4-5, 6-7 and 8-9 directions and $k=4$ corresponding to the D6-brane along 4-5-6-7-8-9
directions. 
Besides these fields, for every pair of D-branes labelled by
$(k,\ell)$ we have two 
chiral superfields $Z^{(k\ell)}$ and
$Z^{(\ell k)}$ arising from open strings stretched between the two D-branes. They 
carry respectively 1 and $-1$ units of charge
under the vector superfield $V^{(k)}$ and $-1$ and 1 units of
charge under the vector superfield $V^{(\ell)}$. 

We shall now write down the action involving these fields. To begin with we shall assume that the
six circles of $T^6$ are orthonormal to each other, with each circle having radius 
$\sqrt{\alpha'}$ and that there is no background 2-form field along $T^6$. 
From now on, we shall set $\alpha'=1$.
In this case the action
takes the form
\be 
S_{kinetic} + \int dx^0 \left[ \int d^4\theta \sum_{k=1}^4 \sum_{\ell =1 \atop \ell\ne k}^4 
\left\{ \bar Z^{(k\ell)} e^{2V^{(\ell)} - 2V^{(k)}} Z^{(k\ell)} \right\}
+ \int d^2 \theta \, \WW + \int d^2\bar \theta \, \overline{\WW} \right]\, ,
\ee
where $S_{kinetic}$ denotes the kinetic terms for the vector 
superfields $V^{(k)}$ and the gauge neutral chiral superfields $\Phi^{(k)}_i$. These
have the standard form and will not be written down explicitly. The superpotential
$\WW$ has two different components. The first component describes the coupling of the 
superfields $\Phi^{(k)}$ to $Z^{(k\ell)}$ and takes the form
\be \label{ew1}
\WW_1 = \sqrt 2 \left[\sum_{k,\ell,m=1}^3 \ve^{k\ell m} \, \Phi^{(k)}_m 
\, Z^{(k\ell)} Z^{(\ell k)} + \sum_{k=1}^3 \Big(\Phi^{(k)}_k - \Phi^{(4)}_k\Big) Z^{(4k)} Z^{(k 4)} \right]
\, ,
\ee
where $\ve^{k\ell m}$ is the totally antisymmetric symbol with $\ve^{123}=1$. The second component
describes the cubic self-coupling between the $Z^{(k\ell)}$'s and takes the form
\be \label{ew2}
\WW_2 = \sqrt 2 \, C\, \sum_{k,\ell, m=1\atop k<\ell,m; \, \ell\ne m }^4 Z^{(k\ell)} Z^{(\ell m)} Z^{(m k)}\, ,
\ee
where $C$ is a constant whose value can be computed in principle by analyzing the coupling
between open strings stretched between different branes, but we shall not need it for
our analysis. 
The sum over $k,\ell, m$ runs over all distinct values of $k$, $\ell$ and $m$ 
which are not related
by cyclic permutations of $(k,\ell,m)$.
There could also be gauge invariant quartic
and higher order terms involving the $Z^{(k\ell)}$'s, but as we shall see, these can be ignored
in our analysis.
 
So far we have assumed that background metric along $T^6$ is diagonal and that there are no
background 2-form fields. We shall now study the effect to switching on small background 
values of the
off-diagonal components of the metric and 2-form fields. As reviewed in appendix
\ref{sa}, this has two
effects. First it introduces Fayet-Iliopoulos (FI) term with coefficient
$c^{(k)}$ for each of the four vector superfields, satisfying
\be \label{eck0}
\sum_{k=1}^4 c^{(k)} = 0\, .
\ee
Second, it generates a linear term in the superpotential of the form
\be \label{ew3}
\WW_3 = \sqrt 2\left[
\sum_{k,\ell,m=1}^3 c^{(k\ell)} \, \ve^{k\ell m} \Phi^{(k)}_m
+ \sum_{k=1}^3 c^{(k4)} \, \Big(\Phi^{(k)}_k - \Phi^{(4)}_k\Big) \right], \quad
c^{(\ell k)} = c^{(k\ell)} \quad \hbox{for} \quad 1\le k<\ell\le 4\, .
\ee
Explicit expressions for $c^{(k)}$ and $c^{(k\ell)}$ for $1\le k <\ell \le 4$ in terms of the off-diagonal
components of the metric and 2-form fields have also been given in appendix \ref{sa}.

Let us now write down the potential involving the scalar fields derived from the above action.
This consists of three pieces. The first comes from the usual quartic coupling between the gauge
field components $X^{(k)}_i$ and the charged scalars $Z^{(k\ell)}$ and takes the form
\be \label{egauge}
V_{gauge} = \sum_{i=1}^3  \sum_{k=1}^4 \sum_{\ell =1\atop \ell \ne k}^4 \,
(X^{(k)}_i - X^{(\ell)}_i) (X^{(k)}_i - X^{(\ell)}_i) 
\Big( \bar Z^{(k\ell)} Z^{(k\ell)} + \bar Z^{(\ell k)} Z^{(\ell k)} \Big)\, ,
\ee
where `bar' denotes complex conjugation.  
The second component of the potential is the D-term contribution. This takes the form
\be \label{evd}
V_D = {1\over 2} \, 
\sum_{k=1}^4 \Big\{ \sum_{\ell=1\atop \ell \ne k}^4 \Big(\bar Z^{(k\ell)} Z^{(k\ell)}
- \bar Z^{(\ell k)} Z^{(\ell k)} \Big) - c^{(k)}\Big\}^2\, .
\ee
The third component is the F-term potential which takes the form
\be \label{evf}
V_F=\sum_{k=1}^4 \sum_{i=1}^3 \left| {\p W\over \p \Phi^{(k)}_i} \right|^2
+ \sum_{k=1}^4   \sum_{\ell=1\atop \ell \ne k}^4 \left| {\p W\over \p Z^{(k\ell)} }\right|^2\, .
\ee
For finding a supersymmetric configuration we have to look for configurations with vanishing
potential. Since the potential is a sum of squares, this requires setting each of these
terms to zero. In \S\ref{s3} we shall look for solutions to these conditions.

Note that the  potential has the following flat directions
\ben \label{eflat}
&& \Phi^{(k)}_m \to \Phi^{(k)}_m+\xi_m, \quad \hbox{for} \quad 1\le k\le 3, \quad k \ne m\, ;
\quad  1\le m\le 3, 
\nonumber \\
&& \Phi^{(k)}_k \to \Phi^{(k)}_k + \zeta_k, \quad \Phi^{(4)}_k \to
\Phi^{(4)}_k+\zeta_k, \quad \hbox{for} \quad 1\le k\le 3\, , \nonumber \\
&& X^{(k)}_i \to X^{(k)}_i + a_i \, ,\quad \hbox{for} \quad 1\le k\le 4, \quad 1\le i\le 3\, ,
\een
where $\xi_m$ and $\zeta_k$ are arbitrary complex numbers and $a_i$ are arbitrary 
real numbers. The $a_i$'s represent overall translation of the system along the non-compact
directions. The symmetries generated by $\xi_m$ and $\zeta_k$ imply
that the potential has six complex flat directions.\footnote{These directions are
all compact since they are associated with translations along $T^6$ and the dual
torus $\wt T^6$.
Thus the quantization of the zero modes associated with 
these bosonic flat directions does not cause any problem and
gives a unique zero energy ground state.}
This corresponds to six exactly massless chiral multiplets. Since each chiral multiplet
contains a Weyl fermion in 3+1 dimensions which has four real components, we have
altogether $6\times 4=24$ real fermion zero modes after dimensional reduction to 0+1
dimensions. The vector superfield 
$\sum_{k=1}^4 V^{(k)}$ also decouples from the action, reflecting the symmetry parametrized
by the $a_i$'s. The Majorana fermion belonging
to this multiplet gives 4 more fermion zero modes. Thus altogether we have
$24+4=28$ fermion zero modes.
These are the Goldstino modes associated with supersymmetry breaking; since
a 1/8 BPS black hole in $\NN=8$ supersymmetric string theory preserves 4 out of
32 supersymmetries, we expect $32-4=28$ broken supersymmetries.
Quantization of these 28
fermion zero modes
gives the $2^{14}$ fold degenerate supermultiplet which is the right degeneracy for a 1/8 BPS
state in a theory with 32 supersymmetries.

\sectiono{Supersymmetric solutions} \label{s3}

We shall now look for configurations preserving supersymmetry, i.e.\ configurations which
make the potential vanish. As noted below \refb{evf}, this requires setting each term in
$V_{gauge}$, $V_D$ and $V_F$ to zero. Furthermore, due to the $U(1)^4$ gauge symmetry of
the original theory, we need to classify solutions up to equivalence relations under these
$U(1)$ gauge symmetries:
\be
Z^{(k\ell)} \to \exp\left[i\left(\theta^{(k)}-\theta^{(\ell)}\right)\right] Z^{(k\ell)}\, ,
\ee
where $\theta^{(k)}$ for $1\le k\le 4$
are the gauge transformation parameters. Note that the overall $U(1)$ --
obtained by setting all the $\theta^{(k)}$'s equal -- acts trivially on the $Z^{(k\ell)}$'s.
Furthermore, since we have fixed $A^{(k)}_0=0$ gauge, we need to demand equivalence 
only under the subgroup of the gauge group that preserves this gauge condition.
This leaves us with the global part of the gauge group,
labelled by time independent $\theta^{(k)}$'s.

We begin by examining the equations $\p \WW/ \p \Phi^{(k)}_i=0$ for $1\le k\le 4$ and
$1\le i\le 3$. Using \refb{ew1}, \refb{ew3} we see that this gives
\be \label{eqzkl}
Z^{(k\ell)} Z^{(\ell k)} = - c^{(k\ell)} \quad \hbox{for} \quad 1\le k< \ell\le 4 \, .
\ee
It follows from this that as long as the $c^{(k\ell)}$ are non-zero for every $k,\ell$
in the range $1\le k<\ell\le 4$, 
none of the $Z^{(k\ell)}$'s can vanish. Eq.\refb{egauge} now gives
\be
X^{(k)}_i=0 \quad \hbox{for} \quad 1\le k\le 4, \quad 1\le i\le 3\, ,
\ee
up to the translation symmetry parametrized by the constants $a_i$ in eq.\refb{eflat}.
Next we consider the $\p\WW /\p Z^{(k\ell)}=0$ equations. This gives
\ben  \label{ezeq}
&& \sum_{m=1}^3 \ve^{k\ell m} \Big(\Phi^{(k)}_m - \Phi^{(\ell)}_m\Big)
\,  Z^{(\ell k)} + C\, \sum_{m=1\atop m\ne k,\ell}^4 Z^{(\ell m)} Z^{(mk)} = 0 \quad \hbox{for} \quad
1\le k, \ell \le 3\, , \nonumber \\
&& \Big(\Phi^{(k)}_k - \Phi^{(4)}_k\Big) Z^{(k 4)} 
+ C\, \sum_{\ell=1\atop \ell \ne k}^3 Z^{(k\ell)} Z^{(\ell 4)} = 0\quad \hbox{for} \quad
1\le k \le 3\, ,\nonumber \\
&& \Big(\Phi^{(k)}_k - \Phi^{(4)}_k\Big)  Z^{(4 k)} 
+ C\, \sum_{m=1\atop m \ne k}^3  Z^{(4m)} Z^{(mk)} = 0 \quad \hbox{for} \quad
1\le k \le 3\, .
\een
These equations serve two purposes. First they determine the combinations
\be
\Phi^{(k)}_m - \Phi^{(\ell)}_m \quad \hbox{for} \quad 1\le k,\ell, m\le 3, \quad k,l,m\, \,
\,  \hbox{distinct},
\qquad \hbox{and} \qquad \Phi^{(k)}_k - \Phi^{(4)}_k \quad \hbox{for} \quad 1\le k\le 3\, ,
\ee
in terms of the $Z^{(k\ell)}$'s. 
This gives 6 linear combinations of the 12 complex scalars $\Phi^{(k)}_i$. The rest 
of the $\Phi^{(k)}_i$'s are associated with flat directions and hence remain undetermined.
Second they give the following relations among the $Z^{(k\ell)}$'s:
\ben \label{ezref}
&& Z^{(k\ell )}  \sum_{m=1\atop m\ne k,\ell}^4 Z^{(\ell m)} Z^{(mk)} 
= Z^{(\ell k)}  \sum_{m=1\atop m\ne k,\ell}^4 Z^{(k m)} Z^{(m\ell)} \quad \hbox{for}
\quad 1\le k,\ell\le 3\, , \nonumber \\
&&  Z^{(4 k)} \, \sum_{\ell=1\atop \ell \ne k}^3 Z^{(k\ell)} Z^{(\ell 4)}  
= Z^{(k 4)}  \sum_{m=1\atop m \ne k}^3  Z^{(4m)} Z^{(mk)} 
 \quad \hbox{for}
\quad 1\le k \le 3\, .
\een

Finally let us turn to the D-term constraints. It is well known that the effect of the D-term
constraints together with quotienting by the $U(1)$ gauge groups is to convert the space
spanned by the coordinates $Z^{(k\ell)}$ to a toric variety. This is parametrized by the
coordinates $Z^{(k\ell)}$ modded out by the complexified $U(1)$ gauge groups {\it after
removing appropriate submanifolds of complex codimension $\ge 1$ from the space spanned
by the $Z^{(k\ell)}$'s}. These submanifolds are obtained by setting one or more $Z^{(k\ell)}$'s
to zero, and depend on the FI parameters $c^{(k)}$. However, since we have seen that the
F-term constraints force all the $Z^{(k\ell)}$'s to be non-zero, removal of these complex
submanifolds has no effect on the final solutions.\footnote{Put another way,
for a generic toric variety, if some equation is given in terms of
homogeneous coordinates, it may have solutions in more than one patch.
Thus, when we translate the equations in terms of coordinates of any single
patch (which does not cover the whole variety) and look for the solutions,
we always have the risk of not having all the solutions.
Fortunately this is not the case here. If we closely look into
what are the regions that are not covered by an arbitrary single patch, we
see that these are the regions where some of the coordinates vanish.
But our $Z^{(k\ell)}$'s cannot vanish
due to the constraint $Z^{(k\ell)}Z^{(\ell k)} = m_{k\ell}$. Thus, although such
regions exist in the toric variety, they are not part of the solution of
our equations. Hence it is enough to work in a single patch only, which is what we
do.
}
Thus, we can proceed by 
parametrizing the variety by an appropriate set of gauge invariant polynomials and forget
about the D-term constraints. Since to start with there are $4\times 3=12$ 
independent $Z^{(k\ell)}$'s,
and we quotient by a $U(1)^3$ gauge group -- the overall U(1) having trivial action on all
the $Z^{(k\ell)}$'s -- we need $12-3=9$ independent gauge invariant coordinates. We take them
to be 
\begin{equation}
\label{polynomial-ring-basis-Abelian-vacuum}
\begin{aligned}
u_1  &\equiv Z^{(12)} Z^{(21)} \, , 
&u_2&\equiv Z^{(23)} Z^{(32)} \, ,
&u_3 &\equiv Z^{(31)} Z^{(13)} \, , \\[1ex]
u_4 &\equiv Z^{(14)} Z^{(41)}  \, ,
&u_5 &\equiv Z^{(24)} Z^{(42)} \, ,
&u_6 &\equiv Z^{(34)} Z^{(43)}  \, , \\[1ex]
u_7  &\equiv Z^{(12)} Z^{(24)} Z^{(41)} \, ,
&u_8  &\equiv Z^{(13)}  Z^{(34)}  Z^{(41)} \, ,
&u_9 &\equiv Z^{(23)}  Z^{(34)}  Z^{(42)}   \, .
\end{aligned}
\end{equation}

We now note that \refb{eqzkl} fixes $u_1,\ldots, u_6$ completely. Thus, the only remaining
variables are $u_7,u_8,u_9$ and the equations to be solved are given in
\refb{ezref}. These actually give three independent equations
\ben \label{ezfin}
    Z^{(23)}   Z^{(31)}  Z^{(12)} +    Z^{(23)}   Z^{(34)}  Z^{(42)} &=& Z^{(32)}   Z^{(21)}   Z^{(13)} 
 +   Z^{(32)}   Z^{(24)}   Z^{(43)} 
   \, , \nonumber \\
  Z^{(24)}  Z^{(41)}  Z^{(12)} + Z^{(24)}  Z^{(43)}  Z^{(32)}  &=&   Z^{(42)}  Z^{(21)}  Z^{(14)} 
+  Z^{(42)}   Z^{(23)}  Z^{(34)} 
   \, , \nonumber \\
 Z^{(34)}    Z^{(41)}   Z^{(13)} + Z^{(34)} Z^{(42)}   Z^{(23)} &=&   Z^{(43)}  Z^{(31)}  Z^{(14)} 
+   Z^{(43)}  Z^{(32)}  Z^{(24)} 
  \, .        
\een
Defining
\be
m_{k\ell} = m_{\ell k} = - c^{(k\ell)} \quad \hbox{for} \quad 1\le k<\ell \le 4\, 
\ee
and using the solutions for $u_1,\ldots, u_6$ given in \refb{eqzkl}, 
eqs.\refb{ezfin} can be expressed as
\ben
\label{u_7-u_8-u_9-system}
u_7 \, u_8^{-1}  &=& m_{24}     \left(  \frac{ m_{24} \, m_{23} \, m_{12} \, u_9^{-1}  - u_7 \, u_8^{-1} \,
 u_9  } { m_{31} \,  u_7 \, u_8^{-1} \, u_9 - m_{23} \,  m_{24}^{2} \, m_{34}  \, u_9^{-1}} \right),
\nonumber \\
u_7 \,u_9  &=& m_{24}     \left( \frac{ m_{12} \, m_{14} \,  u_9  - m_{34} \, m_{23} \, u_7  } { u_7 - u_9 } 
 \right), \nonumber \\
u_8 \,u_9^{-1}  &=& \left( \frac{ m_{34} \,m_{31} \,m_{14}- u_8 \, u_9 }{ u_8 \,  u_9- m_{34} \, m_{23} \,m_{24}  
} \right),           
\een
respectively.
The solutions to the system
\eqref{u_7-u_8-u_9-system}
are given in Table \ref{u_7-u_8-u_9-solutions} of appendix \ref{sd}.
The important point to note from
Table 
\ref{u_7-u_8-u_9-solutions} is that there are 12 distinct solutions. This shows that there
are 12 supersymmetric ground states, in perfect agreement with the prediction 
\refb{eb14} from the dual description. Furthermore since the moduli space of solutions is
zero dimensional, all the solutions carry zero angular momentum after factoring out the
contribution of the goldstino fermion zero modes. This is in agreement with the prediction
from the black hole side.

It is clear from the form of the potential as well as the solutions given in
Table~\ref{u_7-u_8-u_9-solutions} that under a uniform scaling of all the $c^{(k)}$'s and
$c^{(k\ell)}$'s by a real parameter
$\lambda$, the $Z^{(k\ell)}$'s and $\Phi^{(k)}_m$'s 
at the solution (except the ones associated with flat directions) 
scale as $\lambda^{1/2}$. Thus by taking
$\lambda$ to be small we can ensure that each $Z^{(k\ell)}$ and $\Phi^{(k)}_m$
at the solution is small.
In this case the contributions from the
quartic and higher order terms in the superpotential are small compared to the cubic terms
that we have included. This justifies our ignoring such terms for studyng these solutions.
This also justifies our ignoring the fact that $\Phi^{(k)}_i - \Phi^{(\ell)}_i$ for
$1\le k,\ell\le 3$ and $\Phi^{(k)}_k - \Phi^{(4)}_k$ for $1\le k\le 3$ are periodic variables
while solving the eqs.~\refb{ezeq}.

Note however that we have not ruled out existence of solutions where 
$\Phi^{(k)}_i - \Phi^{(k)}_j$ and $Z^{(k\ell)}$'s are of order unity measured in the string
scale. In such cases we must
take into account possible higher order terms in the superpotential, and must also 
include the effect of $\Phi^{(k)}_i$'s being periodic variables so that we have to include
in our analysis also open string states which wind around the various circles on their way
from one D-brane to another.
In other words, full stringy dynamics is needed
for examining the existence of these states. Our experience with BPS state counting
tells us however that the BPS states arise only from low energy fluctuations on the branes
and hence it seems unlikely that there will be new BPS states from the stringy configurations
of the type described above.

\sectiono{Non-abelian generalization} \label{sb}

\def\Tr{{~ \rm Tr}}

In this section we shall generalize the analysis of \S\ref{s2} to the case 
where some of the stacks have more than one brane, i.e. the $N_i$'s introduced in
\S\ref{ssystem} are not all equal to 1. We shall focus on the scalar fields and their
potential since this is what is needed for the counting of supersymmetric solutions.

We begin with a discussion of how the scalar degrees of freedom change in this case.
First of all, the complex scalars $\Phi^{(k)}_i$ and the real scalars $X^{(k)}_i$
become $N_k\times N_k$ hermitian matrices transforming
in the adjoint representation of $U(N_k)$. On the other hand, the complex scalar
$Z^{(k\ell)}$ becomes $N_k\times N_\ell$ complex matrix transforming in the
$(N_k,\bar N_\ell)$ representation of $U(N_k)\times U(N_\ell)$.

Let us now describe the modification of the potential. The superpotential $\WW_1$ given in
\refb{ew1} is generalized to
\ben \label{ew1gen}
\WW_1 &=& \sqrt 2\left[\sum_{k,\ell,m=1}^3 \ve^{k\ell m} \Tr \, \Big(\Phi^{(k)}_m
\, Z^{(k\ell)} Z^{(\ell k)} \Big) + \sum_{k=1}^3 \Tr \, \Big(  Z^{(4k)} \Phi^{(k)}_k Z^{(k 4)} \Big)\right.
\nonumber \\ && \left.
\qquad - \sum_{k=1}^3 \Tr \, \Big(   \Phi^{(4)}_k Z^{(4k)} Z^{(k 4)} \Big)\right]
\, .
\een
The generalization of \refb{ew2} takes the form
\be \label{ew2gen}
\WW_2 = \sqrt 2\, C \, 
\left[\sum_{k,\ell, m=1\atop k<\ell,m; \, \ell\ne m}^4 \Tr \Big(Z^{(k\ell)} Z^{(\ell m)} Z^{(m k)}
\Big)\right]\, .
\ee
The generalization of \refb{ew3} is
\be \label{ew3gen}
\WW_3 =\sqrt 2\left[\sum_{k,\ell,m=1}^3 c^{(k\ell)} \, \ve^{k\ell m} \, N_\ell
\Tr \, \Big(\Phi^{(k)}_m \Big) 
+ \sum_{k=1}^3 c^{(k4)} \, \Big[ N_4
\Tr \, \Big(\Phi^{(k)}_k\Big)  - N_k  \Tr\Big( \Phi^{(4)}_k\Big) \Big]\right]\, .
\ee
There is also an additional superpotential
\be \label{ew4gen}
\WW_4 = -\sqrt 2 \sum_{k=1}^4 \Tr \Big( \Phi^{(k)}_1 \left[\Phi^{(k)}_2, \Phi^{(k)}_3\right] \Big)\, .
\ee
\refb{egauge} generalizes to
\ben \label{egaugegen}
V_{gauge} &=&\sum_{k=1}^4 \sum_{\ell =1\atop \ell \ne k}^4  \sum_{i=1}^3   \,
\Tr \Big[\Big( X^{(k)}_i  Z^{(k\ell)} - Z^{( k\ell)} 
X^{(\ell)}_i \Big)^\dagger  \Big( X^{(k)}_i  Z^{(k\ell)} - Z^{( k\ell)} 
X^{(\ell)}_i \Big)\Big] \nonumber \\
&& + \sum_{k=1}^4 \sum_{i,j=1}^3 \Tr \Big(\big[X^{(k)}_i, \Phi^{(k)}_j\big]^\dagger
\big[X^{(k)}_i, \Phi^{(k)}_j\big]\Big) + {1\over 4} \sum_{k=1}^4 \sum_{i,j=1}^3 
\Tr \Big( [X^{(k)}_i, X^{(k)}_j]^\dagger [X^{(k)}_i, X^{(k)}_j]\Big)\, . \nonumber \\
\een
Finally, the D-term potential \refb{evd} is generalized to
\be \label{evdgen}
V_D = {1\over 2} \, 
\sum_{k=1}^4   \Tr \bigg[\Big( \sum_{\ell=1\atop \ell \ne k}^4 Z^{(k\ell)} Z^{(k\ell)\dagger} 
- \sum_{\ell=1\atop \ell \ne k}^4 Z^{(\ell k)\dagger}  Z^{(\ell k)}  + 
\sum_{i=1}^3  [\Phi^{(k)}_i, \Phi^{(k)\dagger}_i] -  c^{(k)} I_{N_k} \Big)^2 \bigg]\, ,
\ee
where $I_{N_k}$ denotes $N_k\times N_k$ identity matrix. The FI parameters
$c^{(k)}$ now satisfy
\be \label{ecknonabelian}
\sum_{k=1}^4 c^{(k)} N_k = 0\, .
\ee
The coefficients $c^{(k\ell)}$ and $c^{(k)}$ 
are to be chosen so that they reproduce the masses of the $Z^{(k\ell)}$'s correctly.
The equations take the form of \refb{eckl1} and \refb{eckl2} with identical right hand sides,
but the left hand sides are different since the masses of $Z^{(k\ell)}$'s expressed in terms
of $c^{(k)}$'s and $c^{(k\ell)}$'s have additional dependence on the $N_k$'s.

The potential given above has a shift symmetry generalizing \refb{eflat}
\ben \label{eflatgen}
&& \Phi^{(k)}_m \to \Phi^{(k)}_m+\xi_m I_{N_k},  \quad \hbox{for} 
\quad 1\le k \le 3, \quad k \ne m; \quad  1\le m\le 3, 
\nonumber \\
&& \Phi^{(k)}_k \to \Phi^{(k)}_k + \zeta_k I_{N_k}, \quad \Phi^{(4)}_k \to
\Phi^{(4)}_k+\zeta_k I_{N_4}, \quad \hbox{for} \quad 1\le k\le 3\, , \nonumber \\
&& X^{(k)}_i \to X^{(k)}_i + a_i \, I_{N_k}\, , \quad \hbox{for} \quad 1\le i\le 3\, .
\een
This generates six complex translations along compact directions and three real translations
along the non-compact directions.

The $\p\WW/\p\Phi^{(k)}_m=0$ equations give
\ben \label{eqzklgen}
Z^{(k\ell)}  Z^{(\ell k)} &=&  
 -c^{(k\ell)} \, N_\ell  I_{N_k} +[\Phi^{(k)}_k, \Phi^{(k)}_\ell]
\quad \hbox{for} \quad 1\le k, \ell \le 3\, ,  \nonumber \\
Z^{(k4)}Z^{(4k)} &=& - c^{(k4)} \, N_4 I_{N_k}  + \sum_{\ell, m=1}^3 \ve^{k\ell m} \Phi^{(k)}_\ell \Phi^{(k)}_m\, ,
\quad 1\le k \le 3\, , \nonumber \\
Z^{(4k)}Z^{(k4)} &=& - c^{(k4)} \, N_k  I_{N_4} - \sum_{\ell, m=1}^3 \ve^{k\ell m} \Phi^{(4)}_\ell \Phi^{(4)}_m\, ,
\quad 1\le k \le 3\, ,
\een
generalizing \refb{eqzkl}. The $\p\WW/\p Z^{(k\ell)}$ equations give
 \ben \label{ezeqgen}
&& \sum_{m=1}^3 \ve^{k\ell m} \Big( Z^{(\ell k)} \, \Phi^{(k)}_m -\Phi^{(\ell)}_m \, Z^{(\ell k)}
\Big)
+ C\, \sum_{m=1\atop m\ne k,\ell}^4 Z^{(\ell m)} Z^{(mk)} = 0 \quad \hbox{for} \quad
1\le k, \ell \le 3\, , \nonumber \\
&& \Big(\Phi^{(k)}_k  Z^{(k 4)}  - Z^{(k 4)} \Phi^{(4)}_k\Big)
+ C\, \sum_{\ell=1\atop \ell \ne k}^3 Z^{(k\ell)} Z^{(\ell 4)} = 0 \quad \hbox{for} \quad
1\le k \le 3\, ,
\nonumber \\
&& \Big(Z^{(4 k)} \, \Phi^{(k)}_k    - \Phi^{(4)}_k \, Z^{(4 k)} \Big) 
+ C\, \sum_{m=1\atop m \ne k}^3  Z^{(4m)} Z^{(mk)} = 0 \quad \hbox{for} \quad
1\le k \le 3\, ,
\een
generalizing \refb{ezeq}.

It seems reasonable to assume that up to the translation symmetry described in the last
line of \refb{eflatgen}, all the $X^{(k)}_i$'s vanish at the zeroes of the potential
since this makes all the terms in $V_{gauge}$ vanish. 
This will also make the classical bound state have zero size in the non-compact directions.
Furthermore, the effect of D-term constraints
is to take the quotient of the space of solutions to \refb{eqzklgen}, \refb{ezeqgen} by 
complexified $\prod_{k=1}^4 U(N_k)$ gauge transformations. 
Let $\MM$ be the space 
of gauge inequivalent solutions to
 \refb{eqzklgen}, \refb{ezeqgen} after factoring out the zero mode directions
 associated with the shift symmetry \refb{eflatgen}.
The number of supersymmetric states (or more precisely the index $B_{14}$) will be given by the
Euler number of $\MM$. 
Thus, duality symmetry of string theory predicts that
\be
\chi(\MM) = - \wh c(4 N_1 N_2 N_3 N_4)\, ,
\ee
where $\hat c(u)$ has been defined in \refb{ek6.5}.  
If $\MM$ is zero dimensional, then $\chi(\MM)$ just counts the number of solutions as in the 
abelian case. In that case all the microstates would carry strictly zero angular momentum
after factoring out the contribution due to the goldstino fermion modes.

\sectiono{Conclusion} \label{sconc}

In this paper we have set up the general equations whose solutions describe the BPS
states of type II string theory compactified on $T^6$ carrying only RR charges. We have
been able to solve the equations explicitly when the charges take the lowest possible values.
The result is in perfect agreement with the counting of the same states in a U-dual
description.

Admittedly this is only a small beginning of the much more ambitious project. Nevertheless
even at this level our analysis provides a non-trivial test of duality symmetry, since the counting
leading to the magic number 12 is very different from the one that was used to arrive at the
formula \refb{ek6.5}. As far as test of black hole entropy is concerned, a black hole carrying
charges given in \refb{echarge} has large curvature at the horizon and hence the Bekenstein-Hawking
entropy is not expected to agree with $\ln 12$. Nevertheless explicit computation of Bekenstein-Hawking
entropy, together with one loop logarithmic corrections\cite{1005.3044,1106.0080}, 
give a macroscopic entropy
\be
S_{macro} = \pi \sqrt\Delta - 2 \ln \Delta +\cdots \simeq 2\pi - 2 \ln 4 \simeq 3.51\, ,
\ee
which is not very different from the microscopic entropy
\be  \label{emicro}
S_{micro} = \ln 12 = 2.48\, .
\ee
Thus it is not unreasonable to regard our analysis as the counting of microstates of a black hole
made solely of D-branes although the curvature at the horizon of the black hole is large.
Just for comparison we note that for $\Delta=100$, $\ell_1\ell_2=1$ we shall have
\be
S_{macro} = \pi \sqrt{100} - 2 \ln 100 +\cdots \simeq 22.2056, \quad S_{micro} = \ln 3627000060
= 22.012 \, .
\ee

In recent years there has also been progress in computing the macroscopic entropy of 
these black holes by evaluating the supergravity path integral in the near horizon geometry
of the black hole using localization 
techniques\cite{0905.2686,1012.0265,1111.1161,1208.6221,1404.0033}. 
In this approach one regards the $\NN=8$
supersymmetric theory in 3+1 dimensions as an $\NN=2$ supersymmetric theory with 
vector, hyper, gravitino and Weyl multiplets and evaluates the path integral. 
Although the arguments are not complete
due to the inability to extend the analysis to include hypermultiplets and gravitino
multiplets in the language of $\NN=2$ supergravity, 
if we ignore this problem then the result of localization gives the following
result for $S_{macro}$ from the leading saddle point\cite{1111.1161}  
\be
S_{macro} \simeq \ln\left[\sqrt 2\pi \, \Delta^{-7/4} \, I_{7/2}(\pi\sqrt\Delta)\right]\, ,
\ee
where $I_n(x)$ is the standard Bessel function. For $\Delta=4$ this gives
\be
S_{macro}=2.50\, ,
\ee
which is quite close to the microscopic result \refb{emicro}.

Finally we must mention that there is one important aspect of our result which could have
significant impact on our understanding of black hole microstates in the future. 
All the microstates of the D-brane system we have constructed have zero angular momentum
after factoring out the contribution due to fermion zero modes, in agreement
with the prediction from the black hole side. Although the D-brane and black hole descriptions
hold in different regions of the moduli space of the theory, and hence the detailed results
on the angular momentum need not match, the results mentioned above indicate that the
D-brane description may be closer to the actual microstates of the black hole than what one
might naively expect.  This could eventually help us identify the microstates of the black hole
in the region of the moduli space where the black hole description is actually valid.

\bigskip

\noindent {\bf Acknowledgements:} 
We wish to thank Anirban Basu and Boris Pioline
for useful discussions.
This work was
supported in part by the 
DAE project 12-R\&D-HRI-5.02-0303. 
The work of A.S. was also supported in
part by the
J. C. Bose fellowship of 
the Department of Science and Technology, India.

\appendix

\sectiono{Effect of metric and 2-form background} \label{sa}

The undeformed system that we started with in \S\ref{s2} had the six
circles of $T^6$ orthonormal to each other and each having period $2\pi$.
The system of D6-D2-D2-D2 branes that we have considered will be described by
a set of massless degrees of freedom with action governed by the supersymmetric
action with superpotentials $\WW_1$ and $\WW_2$ and all the FI parameters set
to zero. In this appendix we shall show that the effect of small off-diagonal components
of the metric and the 2-form field is to generate the superpotential $\WW_3$ given in
\refb{ew3} and the FI terms labelled by the $c^{(k)}$'s. 

Our strategy will be to compute the mass of the open strings stretched between 
different D-branes in the presence of the deformation and compare it with the
mass computed from the deformed action given in \S\ref{s2}. Consider for example the open
string stretched between the D2-brane along the 4-5 directions and the D6-brane along
the 4-5-6-7-8-9 directions, labelled by the 
0+1 dimensional fields $Z^{(14)}$ and $Z^{(41)}$. The quadratic terms involving
these fields, computed from the action in \S\ref{s2}, 
is given by 
\be \label{equad1}
(c^{(4)}-c^{(1)}) (|Z^{(14)}|^2 - |Z^{(41)}|^2) +4 \,  c^{(14)*} Z^{(14)} Z^{(41)} 
+4 \, c^{(14)} Z^{(14)^*} Z^{(41)*} \, .
\ee
After diagonalization we find the renormalized (mass)$^2$ of the fields to be
\be  \label{elag}
\pm \sqrt{16 |c^{(14)}|^2 + (c^{(1)} - c^{(4)})^2}\, .
\ee
On the other hand, we can calculate the mass of the open string stretched between
the D2-brane along the 4-5 directions and the D6-brane along the 4-5-6-7-8-9 directions 
as follows. First we make a T-duality transformation along
the 4-5-directions to convert this into a D0-D4 system. This leaves unchanged the components
of the metric and 2-form field along the 6-7-8-9 direction. Now for small values of the background
2-form field, the (mass)$^2$ of the
open string stretched between the D0-brane and the
D4-brane along the 6-7-8-9 directions takes the
form\cite{9908142,0007235}
\be \label{edir}
\pm \sqrt{{1\over 2} \sum_{m,n} b_{mn} \left(b^{mn} + {1\over 2} \sum_{p,q}
\epsilon^{mnpq} b_{pq}\right)}
= \pm \sqrt{(b_{67}+ b_{89})^2 + (b_{68} - b_{79})^2 + (b_{69}+b_{78})^2}\, ,
\ee
up to an overall proportionality constant.
Here $\epsilon^{mnpq}$ denotes the 
component of the invariant totally anti-symmetric rank 4
tensor along the D4-brane world-volume. Comparing \refb{elag} and \refb{edir} we get
\be \label{eckl1}
16\, 
|c^{(14)}|^2 + (c^{(1)} - c^{(4)})^2 = (b_{67}+ b_{89})^2 + (b_{68} - b_{79})^2 + (b_{69}+b_{78})^2\, .
\ee
A similar analysis of open strings stretched between other brane pairs, and comparison with the
result derived from the deformed action yields the results
\ben \label{eckl2}
16\, |c^{(12)}|^2 + (c^{(1)} - c^{(2)})^2 &=& (g_{47}+ g_{56})^2 + (b_{45} - b_{67})^2 + (g_{46}-g_{57})^2\, ,
\nonumber \\
16\, |c^{(13)}|^2 + (c^{(1)} - c^{(3)})^2 &=& (g_{49}+ g_{58})^2 + (b_{45} - b_{89})^2 + (g_{48}-g_{59})^2\, ,
\nonumber \\
16\, |c^{(23)}|^2 + (c^{(2)} - c^{(3)})^2 &=& (g_{69}+ g_{78})^2 + (b_{67} - b_{89})^2 + (g_{68}-g_{79})^2\, ,
\nonumber \\
16\, |c^{(24)}|^2 + (c^{(2)} - c^{(4)})^2 &=& (b_{45}+ b_{89})^2 + (b_{48} - b_{59})^2 + (b_{49}+b_{58})^2\, ,
\nonumber \\
16\, |c^{(34)}|^2 + (c^{(3)} - c^{(4)})^2 &=& (b_{45}+ b_{67})^2 + (b_{46} - b_{57})^2 + (b_{47}+b_{56})^2\, .
\een 
Note that
in the mass formula, only 6 independent combinations of metric components and 9
independent combinations of 2-form field components appear. This gives a total 
of 15 independent real
quantities. On the other hand,
in our Lagrangian we have 3 independent FI parameters and six complex parameters
$c^{(k\ell)}$. This also gives a total of 15 real parameters. Nevertheless the solutions are not unique
since, for example, the left hand sides of eqs.~\refb{eckl1}, \refb{eckl2} 
are insensitive to the phases of the
$c^{(k\ell)}$'s. A similar symmetry exists on the right hand side.
A possible choice of $c^{(k)}$'s and $c^{(k\ell)}$'s is
\ben \label{esoldir}
&& c^{(1)} = {1\over2} \left(b_{45}-b_{67}-b_{89}\right), \quad c^{(2)} ={1\over 2} \left(
b_{67} - b_{45} - b_{89}\right), \nonumber \\
&& c^{(3)} = {1\over 2} \left(b_{89}- b_{45} - b_{67}\right), 
\quad
c^{(4)} = {1\over 2} \left(b_{45} + b_{67} + b_{89}\right)  \, , 
\een
\ben\label{esoldir1}
&&16\,  |c^{(12)}|^2 = ( g_{47} +  g_{56})^2 + ( g_{46} -  g _{57})^2 , 
\quad 16\, |c^{(13)}|^2 = ( g_{49} +  g_{58})^2 +  ( g_{48} -  g _{59})^2,
\nonumber \\
&&16\,  |c^{(14)}|^2  = (b_{68} - b_{79})^2 + (b_{69} + b_{78})^2, \quad
16\, |c^{(23)}|^2 = ( g_{69} +  g_{78})^2  + ( g_{68} -  g _{79})^2, \nonumber \\
&&16\,  |c^{(24)}|^2 = (b_{48} - b_{59})^2 + (b_{49} + b_{58})^2 , \quad 
16\, |c^{(34)}|^2  =  (b_{46} - b_{57})^2 + (b_{47} + b_{56})^2  \, .
\een
It is also clear that given a set of $c^{(k)}$'s and $c^{(k\ell)}$'s satisfying $\sum_{k=1}^4 c^{(k)}=0$
it is always possible to find $b_{ij}$'s and $g_{ij}$'s satisfying \refb{esoldir}, \refb{esoldir1}.
For example, if we denote by the subscripts $R$ and $I$ the real and imaginary parts
of $c^{(k\ell)}$, then we can invert \refb{esoldir}, \refb{esoldir1} as
\ben \label{esolrev}
&& b_{45} = {1\over 2} (c^{(1)} - c^{(2)} -c^{(3)}+c^{(4)}), \quad
b_{67} = {1\over 2} (-c^{(1)} + c^{(2)} -c^{(3)}+c^{(4)}), \nonumber \\
&& b_{89} = {1\over 2} (-c^{(1)} - c^{(2)} + c^{(3)}+c^{(4)}), \een
\ben
& g_{47} +  g_{56} = 4\, c^{(12)}_R, \quad & g_{46} -  g _{57} =   4\, c^{(12)}_I, 
\quad g_{49} +  g_{58} =  4\, c^{(13)}_R, \quad  g_{48} -  g _{59} =  4\, c^{(13)}_I, \nonumber \\
& b_{68} - b_{79} = 4\,  c^{(14)}_R, \quad &b_{69} + b_{78} =  4\, c^{(14)}_I, \quad 
g_{69} +  g_{78} =  4\, c^{(23)}_R, \quad g_{68} -  g _{79} =  4\, c^{(23)}_I, \nonumber \\
& b_{48} - b_{59} =  4\, c^{(24)}_R, \quad & b_{49} + b_{58} = 4\,  c^{(24)}_I, \quad
b_{46} - b_{57} =  4\, c^{(34)}_R, \quad b_{47} + b_{56} =  4\, c^{(34)}_I\, .
\een
This establishes that the effect of an arbitrary constant background metric and 2-form
field along $T^6$ can be encoded in an appropriate choice of the constants $c^{(k)}$ and
$c^{(k\ell)}$. On the other hand, a generic deformation of the D-brane world-volume 
theory characterized by the
constants $c^{(k)}$ and $c^{(k\ell)}$ can be produced by suitably choosing the background values
of the metric and 2-form fields.

\sectiono{Duality transformation} \label{sc}

In \cite{0506151}, the counting of  BPS states was done for a system consisting of
$N_1$ KK monopoles associated with the 4-direction, $-N_2$ units of momentum along the
5-direction, $N_3$ D1-branes along the 5-direction, $N_4$ D5-branes along 5-6-7-8-9 directions and
$-N_5$ units of momentum along the 4-direction.\footnote{The actual computation was done
for $N_1=1$ but we shall consider a more general situation.}  
Our goal will be to show that via a series of duality transformations this can be related
to the system introduced in \S\ref{ssystem}.
During this analysis we shall ignore all the signs (which can in principle be determined
by following some specific sign convention, {\it e.g.} the one given in appendix A
of \cite{0708.1270}). 
This way we shall at most miss the signs of the charges carried by the final D -brane 
configurations. However, our analysis of the world-line theory of the D-brane system
is independent of the signs of these charges
as long as the signs are chosen to give a configuration that preserves 4
out of 32 supersymmetries.

Consider the following series of duality transformations:
\begin{itemize}
\item T-duality transformations along the 4 and 5 directions: This gives a configuration of
$N_1$ NS-5-branes along 5-6-7-8-9 directions, $N_2$ fundamental strings along the
5-direction,
$N_3$ D1-branes along the 4-direction, $N_4$ D5-branes along 4-6-7-8-9 directions and
$N_5$ fundamental strings along the 4-direction.
\item T-duality transformation along 8 and 9 directions: This  gives a configuration of
$N_1$ NS-5-branes along 5-6-7-8-9 directions, $N_2$ fundamental strings along  the 5-direction,
$N_3$ D3-branes along 4-8-9 directions, $N_4$ D3-branes along 4-6-7  directions and
$N_5$ fundamental strings along the 4-direction.
\item S-duality: This  gives a configuration of
$N_1$ D5-branes along 5-6-7-8-9 directions, $N_2$ D1-branes along  the 5-direction,
$N_3$ D3-branes along 4-8-9 directions, $N_4$ D3-branes along 4-6-7  directions and
$N_5$ D1-branes along the 4-direction.
\item T-duality along 5, 8 and 9 directions: This  gives a configuration of
$N_1$ D2-branes along 6-7 directions, $N_2$ D2-branes along  8-9 directions,
$N_3$ D2-branes along 4-5 directions, $N_4$ D6-branes along 4-5-6-7-8-9  directions and
$N_5$ D4-branes along the 4-5-8-9 directions.
\item Cyclic permutation of 6-7$\to$4-5$\to$8-9$\to$6-7: This  gives a configuration of
$N_1$ D2-branes along 4-5 directions, $N_2$ D2-branes along  6-7 directions,
$N_3$ D2-branes along 8-9 directions, $N_4$ D6-branes along 4-5-6-7-8-9  directions and
$N_5$ D4-branes along the 6-7-8-9 directions.
\end{itemize}
For $N_5=0$, this reduces to the configuration described in \S\ref{s2}.

\sectiono{Explicit solutions to \refb{u_7-u_8-u_9-system}}  \label{sd}

In this appendix we shall describe the solutions to \refb{u_7-u_8-u_9-system} 
in terms of the quantities
\ben
a &\equiv& m_{12}^2  \left(  m_{31} \, m_{14} \, - m_{23} \, m_{24}   \right)^2
\nonumber \\
&\phantom{\equiv}& - 2 \, m_{12} \, m_{34} 
     \left[
            m_{31}^2
            m_{14} \,
            m_{23} \,
          + m_{31} \,
            m_{24}
            \left(
                   m_{14}^2
                 - 4 \,
                   m_{14} \,
                   m_{23} \,
                 + m_{23}^2    
            \right)
          + m_{14} \,
            m_{23} \,
            m_{24}^2    
    \right]
\nonumber
\\
  &\phantom{\equiv}&
   + m_{34}^2
     \left(
            m_{31} \,
            m_{23} \,
          - m_{14} \,
            m_{24}  
     \right)^2\, , \nonumber \\
b &\equiv& - \:   \frac{   m_{23}^2 \left( m_{31} - m_{24}   \right) \left( m_{12} - m_{34}  \right)}
 { m_{14} - m_{23}  } -  \frac{m_{14} \left( m_{12} \, m_{31}- m_{24} \, m_{34} \right)^2 }
  { \left(m_{31} \, - m_{24} \right) \left(m_{12} \, - m_{34}\right) }
\nonumber \\[1ex]
&\phantom{\equiv \ }& + \frac{
            m_{23}
            \left[
                 - \:
                   m_{12}^2 \,
                   m_{31}
                   \left(
                          m_{31}
                        - 2 \,
                          m_{24}
                   \right)
                 + 2 
                   m_{12} \,
                   m_{31} \,
                   m_{34}
                   \left(
                          m_{31}
                        - 2 \,
                          m_{24}  
                   \right)
                 + m_{24}^2
                   m_{34}^2 \,          
            \right] 
          }
          {
            \left(
                   m_{31}
                 - m_{24}
            \right)
            \left(
                   m_{12}
                 - m_{34}   
            \right)       
          } \, ,             \nonumber \\        
c &\equiv&
     \sqrt{a}
     \left[
          - \:
            \frac{
                   m_{23}
                 }
                 { m_{14} 
                 - m_{23}
                 }
          - \:
            \frac{ m_{12} \, 
                   m_{31}
                 }
                 { 
                   \left(
                          m_{31} 
                        - m_{24}
                   \right)
                   \left(
                          m_{12} 
                        - m_{34}
                   \right)                        
                 }  
          +\:
            \frac{ m_{24} \, 
                   m_{34}
                 }
                 { 
                   \left(
                          m_{31} 
                        - m_{24}
                   \right)
                   \left(
                          m_{12} 
                        - m_{34}
                   \right)                        
                 }
     \right],  \nonumber \\
d &\equiv&
   - \:
     m_{12}
     \left[
            m_{14} \,
            \sqrt{a}
          + 2 \,
            m_{31} \,
            m_{14} \,
            m_{23} \,
            m_{34}
          + m_{24} \,
            m_{34}
            \left(
                   m_{14}^2
                 - 4 \,
                   m_{14} \,
                   m_{23}
                 + m_{23}^2     
            \right)     
     \right]   
\nonumber
\\
  &\phantom{\equiv \ }&
   + m_{23} \,
     m_{34}
     \left(
            \sqrt{a}
          + m_{31} \,
            m_{23} \,
            m_{34}
          - m_{14} \,
            m_{24} \,
            m_{34}   
     \right)
   + m_{12}^2 \,
     m_{14}
     \left(
            m_{31} \,
            m_{14}
          - m_{23} \,
            m_{24}   
     \right),      \nonumber \\
e &\equiv&
   - \sqrt{a}
   + m_{12}
     \left(
            m_{31} \,
            m_{14}
          + m_{23} \,
            m_{24}
     \right)       
   - m_{34}
     \left(
            m_{31} \,
            m_{23}
          + m_{14} \,
            m_{24}  
     \right),      \nonumber \\
e' &\equiv&
   - \sqrt{a}
   - m_{12}
     \left(
            m_{31} \,
            m_{14}
          + m_{23} \,
            m_{24}
     \right)       
   + m_{34}
     \left(
            m_{31} \,
            m_{23}
          + m_{14} \,
            m_{24}  
     \right),      \nonumber \\
f &\equiv&
     \sqrt{a} \,
     m_{12} \,
     m_{14}
   + m_{23} \,
     m_{34}
     \left(
          - \sqrt{a}
          + m_{31} \,
            m_{23} \, 
            m_{34}
          - m_{14} \,
            m_{24} \,
            m_{34}    
     \right)
\nonumber
\\
  &\phantom{\equiv}&   
   + m_{12}^2 \,
     m_{14} 
     \left(
            m_{31} \,
            m_{14}
          - m_{23} \,
            m_{24}  
     \right)
\nonumber
\\
  &\phantom{\equiv}&
   - m_{12} \,
     m_{34}
     \left[
            2 \,
            m_{31} \,
            m_{14} \,
            m_{23}
          + m_{24}
            \left(
                   m_{14}^2
                 - 4 \,
                   m_{14} \,
                   m_{23}
                 + m_{23}^2
            \right)            
     \right],        \nonumber \\ 
g &\equiv&
     2 
     \sqrt{2} \,
     m_{23} \,
     m_{24} \,
     m_{34}
     \left(
            m_{12}
          - m_{34}
     \right)
     \left(
            m_{14}
          - m_{23}   
     \right), \nonumber \\
h &\equiv&
     2 
     \sqrt{2} \,
     m_{23} \,
     m_{24}
     \left(
            m_{12}
          - m_{34}
     \right) \, .
     \een
The solutions are given in Table~\ref{u_7-u_8-u_9-solutions}.
The important point to note is that there are 12 solutions, in agreement with the 
microscopic results.

\begin{table}
\begin{center}
\[
\renewcommand{\arraystretch}{1.5}
\begin{array}{r|c|c|c|}
\cline{2-4}
 & u_7 & u_8 & u_9
\\[1ex]
\cline{2-4}
1 & 
- \sqrt{ m_{12} \, m_{14} \, m_{24} } & 
- \sqrt{ m_{31} \, m_{14} \, m_{34} } & 
- \sqrt{ m_{23} \, m_{24} \, m_{34} }
\\[1ex]
\cline{2-4}
2 & 
  \sqrt{ m_{12} \, m_{14} \, m_{24} } & 
- \sqrt{ m_{31} \, m_{14} \, m_{34} } & 
- \sqrt{ m_{23} \, m_{24} \, m_{34} }
\\[1ex]
\cline{2-4}
3 & 
- \sqrt{ m_{12} \, m_{14} \, m_{24} } & 
  \sqrt{ m_{31} \, m_{14} \, m_{34} } & 
- \sqrt{ m_{23} \, m_{24} \, m_{34} }
\\[1ex]
\cline{2-4}
4 & 
  \sqrt{ m_{12} \, m_{14} \, m_{24} } & 
  \sqrt{ m_{31} \, m_{14} \, m_{34} } & 
- \sqrt{ m_{23} \, m_{24} \, m_{34} }
\\[1ex]
\cline{2-4}
5 &
- \sqrt{ m_{12} \, m_{14} \, m_{24} } & 
- \sqrt{ m_{31} \, m_{14} \, m_{34} } & 
  \sqrt{ m_{23} \, m_{24} \, m_{34} }
\\[1ex]
\cline{2-4}
6 & 
  \sqrt{ m_{12} \, m_{14} \, m_{24} } & 
- \sqrt{ m_{31} \, m_{14} \, m_{34} } & 
  \sqrt{ m_{23} \, m_{24} \, m_{34} }
\\[1ex]
\cline{2-4}
7 & 
- \sqrt{ m_{12} \, m_{14} \, m_{24} } & 
  \sqrt{ m_{31} \, m_{14} \, m_{34} } & 
  \sqrt{ m_{23} \, m_{24} \, m_{34} }
\\[1ex]
\cline{2-4}
8 &
  \sqrt{ m_{12} \, m_{14} \, m_{24} } & 
  \sqrt{ m_{31} \, m_{14} \, m_{34} } & 
  \sqrt{ m_{23} \, m_{24} \, m_{34} }
\\[1ex]
\cline{2-4}
9 &
- (d/g) \sqrt{b+c} & 
  (e/h) \sqrt{b+c} & 
- \sqrt{(b+c)/2}
\\[1ex]
\cline{2-4}
10 & 
  (d/g) \sqrt{b+c} & 
- (e/h) \sqrt{b+c} & 
  \sqrt{(b+c)/2}
\\[1ex]
\cline{2-4}
11 & 
- (f/g) \sqrt{b-c} & 
- (e'/h) \sqrt{b-c} & 
- \sqrt{(b-c)/2}
\\[1ex]
\cline{2-4}
12 &
  (f/g) \sqrt{b-c} & 
  (e'/h) \sqrt{b-c} & 
  \sqrt{(b-c)/2}
\\[1ex]
\cline{2-4}
\end{array}
\]
\end{center}
\caption{Solutions to \eqref{u_7-u_8-u_9-system}. \label{u_7-u_8-u_9-solutions}
}
\end{table}

\vfill \eject

\end{document}